\documentclass[12pt]{revtex4}
\usepackage{graphicx}
\usepackage{amsmath}
\usepackage{amssymb}
\usepackage{hyperref}

\begin{document}

\title{ $B \to A$ transitions  in the light-cone QCD sum rules with the chiral current}
\author{Yan-Jun Sun$^{1,\;2}$\footnote{Email: sunyj@ihep.ac.cn}, Zhi-Gang
Wang$^{3}$, and Tao Huang$^1$ \footnote{Email: huangtao@ihep.ac.cn}}
\address{$^1$Institute of High Energy Physics and Theoretical Physics Center
for Science Facilities,Chinese Academy of Sciences, Beijing 100049, P.R. China\\
$^2$Department of Modern Physics, University of Science and
Technology of China, Hefei 230026, P.R. China\\
$^3$Department of Physics, North China Electric Power University,
Baoding 071003, P. R. China}

\begin{abstract}
In this article, we calculate the form-factors of the transitions $B
\to a_1(1260)$, $b_1(1235) $ in the leading-order approximation
using the  light-cone QCD sum rules. In calculations, we choose the
chiral current to interpolate the $B$-meson, which has outstanding
advantage that the twist-3  light-cone distribution amplitudes of
the axial-vector mesons  have no contributions, and the resulting
sum rules for the form-factors suffer from much less uncertainties.
Then we study the semi-leptonic decays $B \to a_1(1260) l\bar{\nu}_l$,
$b_1(1235) l\bar{\nu}_l$ $(l=e,\mu,\tau)$, and make predictions for
the differential decay widths and decay widths, which can be
confronted with the experimental data in the coming future.
\end{abstract}

\maketitle


\section{Introduction}
The semi-leptonic $B$-decays are excellent subjects in exploring the
CKM matrix elements  and CP violations. We can use both the
exclusive and inclusive $b \to u$ transitions  to study the CKM
matrix element $V_{ub}$. Although   the inclusive decays are
relatively easier  in theoretical studies, the experimental
measurements are very difficult. Furthermore,  the  perturbative QCD
calculations in the region near the end-point of the lepton spectrum
are less reliable as  many resonances appear \cite{Deandrea:1998ww}.
We can resort to the exclusive processes, which are easy to measure
experimentally, to overcome the difficulty, and study the hadronic
matrix elements with  some nonperturbative methods, such as the
light-cone QCD sum rules and lattice QCD.

The relevant exclusive semi-leptonic  decays in determining the CKM
matrix element $V_{ub}$ are $B \to \pi l\bar{\nu}_l$, $ \rho
l\bar{\nu}_l$, $A l\bar{\nu}_l$, where  $A$ denotes the axial-vector
mesons. The semi-leptonic decays $B \to \pi l\bar{\nu}_l$,  $ \rho
l\bar{\nu}_l$, which were firstly observed  by the CLEO
collaboration \cite{Alexander:1996qu},  have been extensively
studied theoretically. The semi-leptonic decays $B \to A
l\bar{\nu}_l$ are expected to be  observed at the LHCb, where the
$b\bar{b}$ pairs will be copiously produced with the cross section
about $500 ~\mu b$ \cite{LHCb}. The  $B \to a_1(1260)$ form-factors
have been studied with the  constituent quark meson (CQM) model
\cite{Deandrea:1998ww}, the covariant light-front (CLF) approach
\cite{Cheng:2003sm}, the improved Isgur-Scora-Grinstein-Wise (ISGW2)
model \cite{Scora:1995ty}, the QCD sum rules (QCDSR)
\cite{Aliev:1999mx}, the light-cone QCD sum rules (LCSR)
\cite{Wang:2008bw,Yang:2008xw} and the perturbative QCD (pQCD)
\cite{Li:2009tx}, and the values differ from each other remarkably.
It is interesting to restudy the semi-leptonic decays $B \to
a_1(1260) l\bar{\nu}_l$, $b_1(1235) l\bar{\nu}_l$ with the chiral
current using the LCSR
\cite{Huang:1998gp,Huang:1999tt,Wang:2002hz,Huang:2001xb,Sun:2010nv}.

In the light-cone QCD sum rules \cite{Balitsky:1989ry}, we carry out the operator product
expansion near the light-cone $x^2\approx 0$ instead of the short
distance $x\approx 0$, while the nonperturbative hadronic matrix
elements  are parameterized by the light-cone distribution
amplitudes (LCDAs) of increasing twist instead of  the vacuum
condensates. Based on the quark-hadron duality, we can obtain
copious information about the hadronic parameters at the
phenomenological side, for example, the form-factors. The twist-2
and twist-3 LCDAs usually enter the sum rules and play an important
role in the LCSR for the form-factors. A better understanding of
those LCDAs is critical to make the calculations more reliable. In
Refs.\cite{Yang:2005gk,Yang:2007zt}, K. C. Yang proposes model LCDAs
for the axial-vector mesons, which are expanded in terms of  the
Gegenbauer polynomials, and estimates the coefficients of the
 LCDAs  with the QCD sum rules. If we choose the
chiral currents, the  twist-3 LCDAs have no contributions to the
form-factors, the uncertainties originate from the LCDAs can be
reduced remarkably
\cite{Huang:1998gp,Huang:1999tt,Wang:2002hz,Huang:2001xb,Sun:2010nv}.
In this article, we extend  our previous works to study  the
semi-leptonic decays $B \to a_1(1260) l\bar{\nu}_l$, $b_1(1235)
l\bar{\nu}_l$.

The paper is organized as follows: In Sec.II, we study the $B \to
a_1(1260)$, $b_1(1235)$ form-factors with the chiral current using
the LCSR; in Sec.III, we present the numerical results of the
form-factors, the differential decay widths and decay widths of the
$B\to a_1(1260) l \bar{\nu}_l$, $b_1(1235) l \bar{\nu}_l$; Sec.IV is
reserved for summary and discussion.

\section{The light-cone sum rules  with the chiral current}

We extend our previous works
\cite{Huang:1998gp,Huang:1999tt,Huang:2001xb,Wang:2002hz,Sun:2010nv}
to study the $B\to A$ form-factors with the chiral current in the
framework of the LCSR. The chiral current warrants  that the LCDAs
of the same (opposite) chirality remain (disappear).

In the standard model, the semi-leptonic decays $B \to A l
\bar{\nu}_l$ take place through the following effective Hamiltonian:
 \begin{eqnarray}
{\cal H}_{eff} &=&\frac{G_F}{\sqrt{2}}V_{ub}\bar
 u\gamma_{\mu}(1-\gamma_5)b \bar l\gamma^{\mu}(1-\gamma_5)\nu_l\, ,\label{eq:Ampbtos} \nonumber
 \end{eqnarray}
where the $V_{ub}$ is the CKM matrix element and the $G_F$ is the
Fermi constant. In calculations, we are confronted with the hadronic
matrix elements $\langle A(P,\epsilon^*)|\bar
{q}\gamma_{\mu}\gamma_5b|\bar{B}(P+q)\rangle$ and $\langle
A(P,\epsilon^*)|\bar {q}\gamma_{\mu} b| \bar{B}(P+q)\rangle$, which
can be  parameterized in terms of the form-factors $A(q^2)$,
$A_1(q^2)$, $A_2(q^2)$, $A_3(q^2)$ and $A_0(q^2)$
\cite{Cheng:2003sm},
 \begin{eqnarray}
  \langle A(P,\epsilon^*)|\bar q\gamma_{\mu}\gamma_5 b|\bar B(P+q)\rangle
   &=&-\epsilon_{\mu\nu\rho\sigma}\epsilon^{*\nu}q^\rho P^\sigma
   \frac{2iA(q^2)}{m_B-m_A} \,  , \label{eq:AA}\\
  \langle A(P,\epsilon^*)|\bar q\gamma_{\mu}b|\bar
  B(P+q)\rangle
   &=&-\epsilon^*_\mu(m_B-m_A)A_1(q^2)+\epsilon^*\cdot q P_{\mu}\frac{2A_{2}(q^2)}{m_B-m_A} \label{eq:Ap}\nonumber \\
   &&+\epsilon^*\cdot q q_{\mu}\left[\frac{A_2(q^2)}{m_B-m_A}+2m_A \frac{A_3(q^2)-A_0(q^2)}{q^2}\right]\, ,
 \end{eqnarray}
where
$A_3(q^2)=\frac{m_B-m_A}{2m_A}A_1(q^2)-\frac{m_B+m_A}{2m_A}A_2(q^2)$,
$A_3(0)=A_0(0)$, $\epsilon^{0123}=1$, and the $\epsilon^{*}_{\nu}$
is the polarization vector of the axial-vector meson. The hadronic
matrix element  $\langle A(P,\epsilon^*)|\bar {q}\gamma_{\mu} b|
\bar{B}(P+q)\rangle$ can be redefined as
 \begin{eqnarray}
  \langle A(P,\epsilon^*)|\bar q\gamma_{\mu}b|\bar
  B(P+q)\rangle
   &=&-\epsilon^*_\mu(m_B-m_A)A_1(q^2) \label{eq:APP}\\
   &&+\epsilon^*\cdot q P_{\mu}\frac{2A_{+}(q^2)}{m_B-m_A}
   +\epsilon^*\cdot q q_{\mu}\frac{A_{+}(q^2)+A_{-}(q^2)}{m_B-m_A} \, ,\nonumber
 \end{eqnarray}
 where
\begin{eqnarray}
A_2(q^2) &=& A_+(q^2)  \, ,\label{eq:A2+}\\
A_3(q^2) &=& \frac{m_B-m_A}{2m_A}A_1(q^2)-\frac{m_B+m_A}{2m_A}A_+(q^2) \, ,\label{eq:A3+}\\
A_0(q^2)&=&\frac{m_B-m_A}{2m_A}A_1(q^2)-\frac{m_B+m_A}{2m_A}A_+(q^2)
           -\frac{q^2}{2m_A(m_B-m_A)}A_-(q^2) \, . \label{eq:A0+}
\end{eqnarray}

In the following,  we write down the  correlation function with a
chiral current,
\begin{eqnarray}
\Pi_{\mu}(P,q)&=& i \int d^4 xe^{iqx} \langle A(P,\perp)|T\left\{
\bar{q_1}(x)\gamma_{\mu}(1-\gamma_5)b(x),\bar{b}(0)i(1+\gamma_5)q_2(0)\right\}
|0 \rangle \, ,\label{eq:pia1}
\end{eqnarray}
where $P^2=m_A^2$. We study  the relevant form-factors with the
transversely polarized  axial-vector mesons \cite{Yang:2008xw},
 and obtain  simple relations among the form-factors as the
 corresponding ones in the  $B\rightarrow V$ transitions.

According to the quark-hadron duality \cite{Shifman:1978bx} and
unitarity, we can insert a complete set of intermediate states with
the same quantum numbers as the current operator
$\bar{b}(0)i(1-\gamma_5)q_1(0)$ in the correlation function to
obtain the hadronic representation. After isolating the ground state
contribution from the pole term of the pseudoscalar $B$ meson, we
obtain the result,
\begin{eqnarray}
\Pi_{\mu} (P,q)&=& \frac{\langle A(P,\perp)|\bar
{q}_1\gamma_{\mu}(1-\gamma_5)b|\bar{B}(P+q)\rangle \langle
\bar{B}(P+q)|\bar{b}i \gamma_5 q_2 |0\rangle}{m^2_B-(P+q)^2}\label{eq:spec}\\
&&+\sum_h \frac{\langle A(p,\perp)|\bar
{q}_1\gamma_{\mu}(1-\gamma_5)b|\bar{B}^h(P+q)\rangle \langle
\bar{B}^h(P+q)|\bar{b}i(1+ \gamma_5) q_2 |0\rangle}{m^2_B-(P+q)^2}\,
. \nonumber
\end{eqnarray}
It should be stressed that there are contributions from the scalar
$B$-meson,  the pseudoscalar $B$-meson, and their resonances
\cite{Chernyak:1990ag}, we can attribute the (ground state) scalar
$B$-meson into  the higher  resonances and continuum states
$|B^h\rangle$. Taking into account the definition of the  $B$-meson
decay constant $\langle \bar{B}|\bar b i\gamma_5 q_2 |0
\rangle=\frac{f_B m_B^2 }{m_{q_2}+m_b}$, we can obtain the hadronic
representation,
\begin{eqnarray}
\Pi_{\mu} (P,q)&=&
\Big[-(m_B-m_A)A_1\epsilon^*_{\perp\mu} +\left(\frac{A_2(q^2)}{m_B-m_A}+2m_A \frac{A_3(q^2)-A_0(q^2)}{q^2}\right) \epsilon_{\perp}^*\cdot q q_{\mu} \nonumber\\
&&+\frac{2A_{2}(q^2)}{m_B-m_A}\epsilon_{\perp}^*\cdot q P_{\mu}
+\frac{2iA(q^2)}{m_B-m_A}
   \epsilon_{\mu\nu\rho\sigma}\epsilon_{\perp}^{*\nu}q^\rho P^\sigma\Big]
     \frac{1}{m^2_B-(P+q)^2}\frac{m^2_B f_B}{m_{q_2}+m_b} \nonumber \\
 &&+\frac{1}{\pi}\int_{s_0}^{\infty} ds \frac{ \rho^h_{\mu}(s) }{s-(P+q)^2} \,
 .
\end{eqnarray}
The spectral density $\rho^h_{\mu}(s)$ can be approximated as
\begin{eqnarray}
\rho^h_{\mu}(s)&=&\rho^{QCD}_{\mu}(s)\theta(s-s_0)\, ,
\end{eqnarray}
by invoking the quark-hadron duality ansatz, where the
$\rho^{QCD}_{\mu}(s)$ is the perturbative  QCD spectral density.
Here the threshold $s_0$ is near the squared mass of the lowest
scalar B-meson.

Now, we briefly outline the operator product expansion for the
correlation function  in perturbative QCD. The calculations are
performed at the large space-like momentum region
 $(P+q)^2\ll m^2_b$ and $0\leq q^2<(m_b-m_A)^2-2(m_b-m_A)\Lambda_{QCD}$
\cite{Colangelo:2000dp}, or more specific,  $0\leq q^2<12\,
\mbox{GeV}^2$ for the axial-vector mesons $a_1(1260)$ and
$b_1(1235)$. We  contract the $b$-quark fields  in  the correlation
function,   substitute  it  with the free $b$-quark propagator, and
obtain the result,
\begin{eqnarray}
\Pi_{\mu}(P,q) &=&i \int \frac{d^4
kd^4x}{(2\pi)^4}\frac{e^{i(q-k)x}}{m^2_b-k^2} Tr
\left\{\left[\gamma_{\mu}(1-\gamma_5)(\not\!k+m_b)(1+\gamma_5)\right]_{\delta
\alpha} \langle A(P,\perp)|\bar q_{{1 \delta}}(x)q_{2
\alpha}(0) |0 \rangle \right\}. \nonumber \\
\end{eqnarray}
The  light-cone  distribution amplitudes of the axial-vector mesons
are defined by \cite{Yang:2008xw}
 \begin{eqnarray}\label{eq:fourier}
  &&  \langle A(P,\lambda)|\bar q_{1\,\delta}(x) \, q_{2\, \alpha}(0)|0\rangle
= -\frac{i}{4} \, \int_0^1 du \,  e^{i u \, P x}
\nonumber\\[0.1cm]
  && \quad \times\,\Bigg\{ f_A m_A \Bigg[
  \not\! P\gamma_5 \, \frac{\epsilon^*_{(\lambda)} x}{Px} \,
  \Bigg(\Phi_\parallel(u) + \frac{m_A^2 x^2}{16} \mathbf{A}_\parallel^2(u)\Bigg)
  +\Bigg( \not\! \epsilon^* -\not\! P \frac{\epsilon^*_{(\lambda)} z}{Px}\Bigg)\,
   \gamma_5 g_\perp^{(a)}(u) \nonumber\\
  && \qquad - \not\! x\gamma_5 \frac{\epsilon^*_{(\lambda)} x}{2(Px)^2}
  m_A^2 \bar g_3(u) +
 \epsilon_{\mu\nu\rho\sigma} \,
    \epsilon^*_{(\lambda)}{}^\mu  P^{\rho} x^\sigma \, \gamma^\mu
    \, \frac{g_\perp^{(v)}(u)}{4}\Bigg]
\nonumber \\[0.1em]
  && \qquad
  + \,f^{\perp}_A \Bigg[
  \frac{1}{2}\bigg( \! \not\! P\not\!\epsilon^*_{(\lambda)}-
  \not\!\epsilon^*_{(\lambda)} \not\! P  \bigg) \gamma_5
 \Bigg(\Phi_\perp(u) + \frac{m_A^2 x^2}{16} \mathbf{A}_\perp^2(u)\Bigg)
 \nonumber\\
 && \qquad
 \,\,\, -
   \frac{1}{2}\bigg( \! \not\! P\not\! x- \not\! x \not\! P  \bigg)
   \gamma_5 \frac{\epsilon^*_{(\lambda)} x}{(Px)^2} m_A^2 \bar
   h_\parallel^{(t)} (u)
 -\frac{1}{4}\bigg( \! \not\!\epsilon^*_{(\lambda)} \not x-
  \not x \not\!\epsilon^*_{(\lambda)} \bigg) \gamma_5
  \frac{m_A^2 }{Px} \bar h_3 (u)\nonumber\\
 && \qquad\,\,\, +i \Big(\epsilon^*_{(\lambda)} x\Big) m_A^2
 \gamma_5
 \frac{h^{(p)}_\parallel (u)}{2}
  \Bigg]\Bigg\}_{\alpha\delta}\, ,
 \end{eqnarray}
where the $u$ is the  fraction of the light-cone momentum of the
axial-vector meson carried by the quark, and $\bar{u}=1-u$. After
carrying out the integrals of the $x$ and $k$, we obtain the
following result,
\begin{eqnarray}
\Pi_{\mu}(P,q) &=& i \int du
 Tr \Bigg\{[\gamma_{\mu}(1-\gamma_5)(\not\!k+m_b)(1+\gamma_5)]_{\delta
\alpha} M^A_{\perp \alpha \delta }\Bigg\}
\frac{1}{m^2_b-k^2}\Bigg|_{k=q+uP} ,\label{eq:pitr}
\end{eqnarray}
where the transverse
 projectors, which project the transverse
 components of the axial-vector meson,
are given by  \cite{Yang:2008xw},
\begin{eqnarray}
M^A_\perp &=& i\frac{f^{\perp}_A}{4} E \Bigg\{\not\!
\epsilon^{*(\lambda)}_\perp\not\! n_- \gamma_5 \,
   \Phi_\perp(u)\nonumber\\
&&  - \frac{f_A}{f_A^\perp}\frac{m_A}{E} \,\Bigg[ \not\!
\epsilon^{*(\lambda)}_\perp\gamma_5 \, g_\perp^{(a)}(u) -  \,
E\int_0^u dv\, (\Phi_\parallel(v) - g_\perp^{(a)}(v)) \not\!
n_-\gamma_5 \, \epsilon^{*(\lambda)}_{\perp\mu}
\,\frac{\partial}{\partial
         k_{\perp\mu}}
\cr && + \,i \varepsilon_{\mu\nu\rho\sigma} \,
       \gamma^\mu \epsilon_\perp^{*(\lambda)\nu} \,  n_-^\rho
         \left( n_+^\sigma \,{g_\perp^{(v)\prime}(u)\over 8}-
          E\,\frac{g_\perp^{(v)}(u)}{4} \, \frac{\partial}{\partial
         k_\perp{}_\sigma}\right)
 \Bigg]
 \, \Bigg|_{k=uP} + {\cal O}\Bigg(\frac{m_A^2}{E^2}\Bigg)
 \Bigg\}\, ,
\end{eqnarray}
here we have taken  $P^\mu = En^\mu_{-} + m^2_A n^\mu_{+}/4E \approx
        En^\mu_{-}$ and the exactly longitudinal and transverse polarization
vectors of the axial-vector meson, independent of the coordinate
variable $x$, are defined as
\begin{eqnarray}
\epsilon^{*(L) \mu}_{\perp} &=& \frac{E}{m_A}\left [\left
(1-\frac{m^2_A}{4E^2}\right
)n^\mu_{-}-\frac{m^2_A}{4E^2}n^\mu_{+}\right ] \, ,\\
\epsilon^{*(\lambda) \mu}_{\perp} &=& \epsilon^{*(\lambda)\mu}
-\frac{\epsilon^{*(\lambda)}n_{+}}{2}n^\mu_{-}
-\frac{\epsilon^{*(\lambda)}n_-}{2}n^\mu_{+}, \quad (\lambda=\pm)\,.
\end{eqnarray}
We carry out  the trace in  Eq.(\ref{eq:pitr}), and observe that
only the leading-twist LCDAs  $\Phi_\perp(u)$ have  contributions,
\begin{eqnarray}
\Pi_{\mu}(P,q)& = & f^{\bot}_A\int^1_0 du
\frac{\Phi_{\bot}(u)}{m^2_b-(q+uP)^2} \left[ 2P\cdot
(q+uP)\epsilon^*_{\perp\mu}-2(\epsilon^*_{\perp}\cdot q)P_{\mu} - 2i
\epsilon_{\mu\nu\rho\sigma}\epsilon^{*\nu}_{\perp}q^{\rho}P^{\sigma}
\right]\, .\nonumber
\end{eqnarray}

\begin{figure}
\includegraphics[totalheight=5cm,width=6cm]{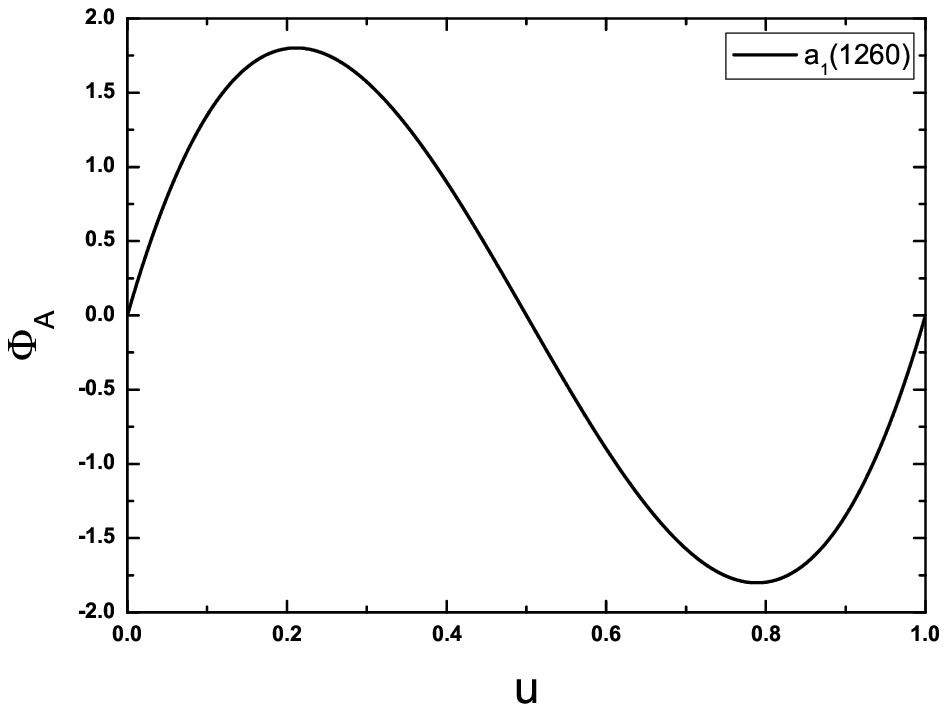}
\includegraphics[totalheight=5cm,width=6cm]{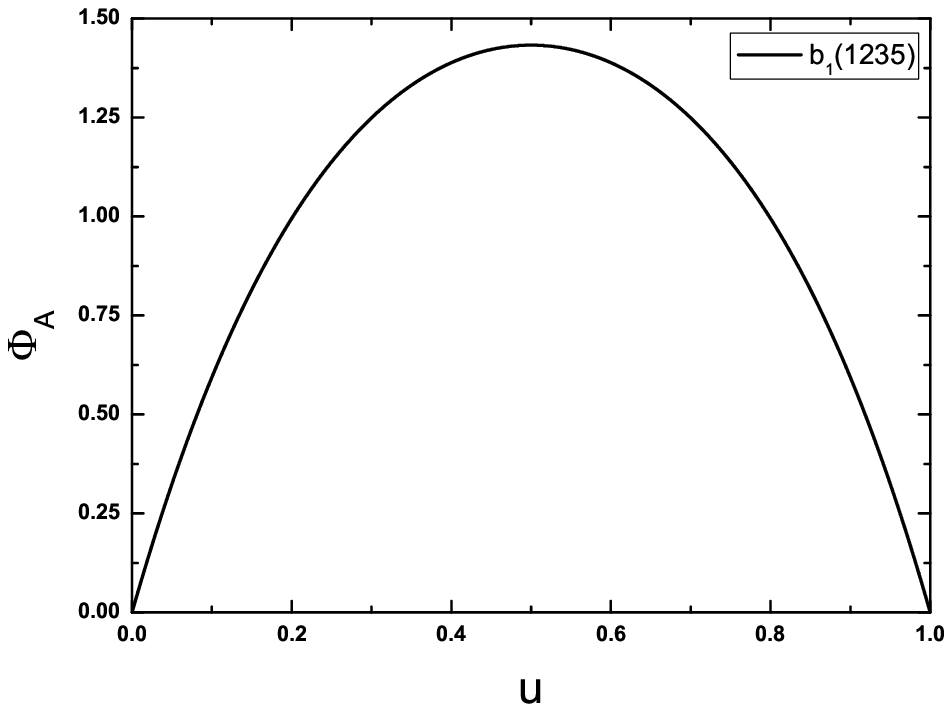}
\caption{The twist-2 LCDAs $\Phi_{\perp}(u,\mu)$ of the axial-vector
mesons $a_1(1260)$ and $b_1(1235)$ at the energy scale $\mu=1.0
~\mbox{GeV}$ \cite{Yang:2005gk,Yang:2007zt}.} \label{figDA}
\end{figure}

With reference to the LCDAs and decay constants of the axial-vector
mesons, a few words should be given. In the flavor $SU(3)$ symmetry
limit, due to G-parity the twist-2 LCDA $\Phi_\perp(u)$  obeys the
normalization
\begin{eqnarray}
\int^1_0 du\Phi_\perp(u)&=&0\label{decay constant1}
\end{eqnarray}
for the  $^3P_1$ meson and
\begin{eqnarray}
\int^1_0 du\Phi_\perp(u) &=&1\label{decay constant2}
\end{eqnarray}
for the $^1P_1$ meson. Based on the conformal symmetry of the QCD
Lagrangian, $\Phi_\perp(u,\mu)$ can be expanded in terms of  a
series of Gegenbauer polynomials $C^{3/2}_{m}(\xi)$ with increasing
conformal spin \cite{Yang:2005gk,Yang:2007zt},
\begin{eqnarray}
\Phi_\perp(u,\mu) & = & 6 u \bar u \left[ a_0^\perp(\mu) + a_1^\perp(\mu)\,C^{3/2}_{1}(\xi) + a_2^\perp(\mu)\,  C^{3/2}_{2}(\xi) + \cdots \right] \label{eq:lcda-3p1-t2-1},
\end{eqnarray}
where $\xi=2u-1$, the values of the coefficients  $a^\perp_m(\mu)$
  at the energy scale $\mu=1 ~\mbox{GeV}$ are
$ a_0^\perp=a_2^\perp=0$, $a_1^\perp=-1.04 \pm 0.34$ for the
$a_1(1260)$ and $a_0^\perp=1$, $a_1^\perp=0$, $a_2^\perp=0.03 \pm
0.19$ for the $b_1(1235)$, respectively. We plot the LCDAs
$\Phi_\perp(u,\mu)$ of the axial-vector mesons $a_1(1260)$ and
$b_1(1235)$   at the energy scale $\mu=1.0~\mbox{GeV}$ in
Fig.\ref{figDA}. The G-parity conserving decay constants of the
axial-vector mesons are defined by \cite{Yang:2008xw}
 \begin{eqnarray}
\langle 1^3P_1(P,\lambda) |\bar q_1\gamma_\mu\gamma_5 q_2 |0\rangle
&=&if_{^3P_1} m_{^3P_1} \epsilon_\mu^{*(\lambda)},  \label{eq:f3p1z} \\
\langle 1^1P_1(P,\lambda) |\bar q_1 \sigma_{\mu\nu}\gamma_5q_2 |0\rangle
 &=&f_{^1P_1}^\perp (\epsilon^{*(\lambda)}_{\mu} P_\nu -
\epsilon^{*(\lambda)}_{\nu} P_\mu),  \label{eq:f1p1h}
 \end{eqnarray}
where the decay constant $f_{^3P_1}$ ($f^\perp_{^1P_1}$) is scale
independent (dependent). The G-parity violating decay constants are
defined by $f^\perp_{^3P_1}=f_{^3P_1}$ and
$f_{^1P_1}=f^\perp_{^1P_1}$ at the energy scale $\mu=1\,\rm{GeV}$.

After matching  with the hadronic representation and performing the
Borel transformation
 with respect to the variable $(P+q)^2$, we
obtain the sum rules for the form-factors:
\begin{eqnarray}
A(q^2) &=&-\frac{m_{q_2}+m_b}{m^2_B f_B}(m_B-m_A)f^{\bot}_A
\int^1_{\Delta} du \frac{\Phi_{\bot}(u)}{u}e^{FF} \label{eq:A} ,\\
A_1(q^2) &=& -\frac{m_{q_2}+m_b}{m^2_B
f_B}\frac{f^{\bot}_A}{m_B-m_A} \int^1_{\Delta} du
\frac{\Phi_{\bot}(u)}{u}\frac{m^2_b-q^2+u^2P^2}{u}e^{FF}
\label{eq:A1} ,\\
A_{2}(q^2) &=&-\frac{m_{q_2}+m_b}{m^2_B f_B}(m_B-m_A)f^{\bot}_A
\int^1_{\Delta} du \frac{\Phi_{\bot}(u)}{u}e^{FF} \label{eq:A2} ,\\
A_3(q^2)&=& -\frac{m_{q_2}+m_b}{m^2_B f_B}\frac{f^{\bot}_A}{2 m_A}
\int^1_{\Delta} du
\frac{\Phi_{\bot}(u)}{u}\frac{m^2_b-q^2+u^2P^2}{u}e^{FF}\nonumber\\
 &&+\frac{m_{q_2}+m_b}{m^2_B f_B}\frac{f^{\bot}_A}{2 m_A}(m_B^2-m_A^2)
\int^1_{\Delta} du \frac{\Phi_{\bot}(u)}{u}e^{FF} ,\label{eq:A3}\\
A_0(q^2)&=& -\frac{m_{q_2}+m_b}{m^2_B f_B}\frac{f^{\bot}_A}{2 m_A}
\int^1_{\Delta} du
\frac{\Phi_{\bot}(u)}{u}\frac{m^2_b-q^2+u^2P^2}{u}e^{FF}\nonumber\\
 &&+\frac{m_{q_2}+m_b}{m^2_B f_B}\frac{f^{\bot}_A}{2 m_A}(m_B^2-m_A^2)
\int^1_{\Delta} du \frac{\Phi_{\bot}(u)}{u}e^{FF} \nonumber\\
 &&+\frac{m_{q_2}+m_b}{m^2_B f_B}\frac{q^2 f^{\bot}_A}{2 m_A}
\int^1_{\Delta} du \frac{\Phi_{\bot}(u)}{u}e^{FF},\label{eq:A0}
\end{eqnarray}
where
\begin{eqnarray}
\Delta &=&\frac{1}{2m_A^2} \left[\sqrt{(s_0-m_A^2+Q^2)^2 +
4(m_b^2+Q^2) m_A^2}-(s_0-m_A^2+Q^2)\right]
\label{eq:delta}\nonumber,\\
FF&=&-\frac{1}{uM^2}\left[m_b^2 +u(1-u)m_A^2+(1-u)Q^2\right]
+\frac{m_B^2}{M^2}  \, ,  \nonumber
\end{eqnarray}
 $M^2$ is the Borel  parameter and  $Q^2=-q^2$.
 The form factors $A_+(q^2)$ and $A_-(q^2)$ can be obtained from the relations
  (\ref{eq:A2+}), (\ref{eq:A3+}) and (\ref{eq:A0+}).

It is surprising that the expressions of the form-factors are very
simple,  and only the leading twist LCDA $\Phi_\perp(u,\mu)$ appears
in the final sum rules. The form-factors $A_+$ and $A_-$ have the
following simple relations,
\begin{eqnarray}
A_{-}(q^2) &=& -A_{+}(q^2) \, , \label{eq:A-+}\\
A(q^2) &=& A_{+}(q^2) \, . \label{eq:A+}
\end{eqnarray}
Similar relations can be obtained for the $B\to V$ form-factors if
we  use the chiral current in the LCSR \cite{Huang:2008sn}. The
simple relations obtained for the $B \to S, V, P, A$ form-factors in
Refs.\cite{Sun:2010nv,Huang:2008sn,Huang:2007kb} and the present
work, up to the hard-exchange corrections, are consistent with the
predictions of the soft collinear effective theory
\cite{Bauer:2000yr}.

\section{numerical results and discussions}
The input parameters for  the semi-leptonic decays $B\to a_1(1260) l
\bar{\nu}_l$,  $ b_1(1235) l \bar{\nu}_l$
 are taken as
\cite{Yang:2005gk,Yang:2007zt,Nakamura:2010zzi,Wang:2008da,Khodjamirian:1998ji}:
\begin{equation}
\begin{array}{ll}
G_F=  1.166 \times 10^{-2} {\rm{GeV}^{-2}}, &
|V_{ub}|=3.96^{+0.09}_{-0.09} \times 10^{-3},
\\
m_u(1 {~\mbox{GeV}})=2.8 ~\mbox{MeV}, &  m_d(1 ~\mbox{GeV})=6.8
~\mbox{MeV},\\
m_b=(4.8 \pm 0.1) ~\mbox{GeV},&
\\
m_{e,\mu}=0 ~\mbox{MeV},&m_{\tau}=1776.82 ~\mbox{MeV} ,
\\
 m_{a_1(1260)}=1.23 \pm 0.06~\mbox{GeV},& m_{b_1(1235)}=1.21 \pm 0.07 ~\mbox{GeV},
\\
 f^\perp_{a_1(1260)}=0.238 \pm 0.010~\mbox{GeV},& f^\perp_{b_1(1235)}=0.180 \pm 0.008 ~\mbox{GeV}.
\\
 m_{B_0}=5.279~\mbox{GeV},& f_{B_0}=(0.19 \pm 0.02) ~\mbox{GeV}.
\label{inputs}
\end{array}
\end{equation}

 We take into account the binding energy difference between the
scalar and pseudoscalar $B$ mesons from the QCD sum rules
  in the heavy quark effective
theory \cite{Huang:1998sa}, and choose the suitable  threshold
parameter $s_0$ to avoid contamination from  the  scalar $B$-meson
\cite{Chernyak:1990ag}, and obtain the value $s_0=(33 \pm
1)~\mbox{GeV}^2$, which is smaller than the ones  used in the
 conventional QCD sum rules to reproduce the
experimental values of the pseudoscalar $B$-meson. Also, it is
possible to determine the threshold parameters in other approaches,
among which the scenario suggested in Ref.\cite{Lucha:2009uy} is
more effective. The  Borel parameter $M^2$ shared  by all the QCD
sum rules in the pseudoscalar channel is $M^2=(10-15)~\mbox{GeV}^2$.
 In this interval, the higher  resonances and continuum states contribute less
than $20\%$ and the uncertainties originated from the Borel
parameter $M^2$ are about $(0.7\sim 1.5)\%$.

\begin{figure}
\includegraphics[totalheight=5cm,width=6cm]{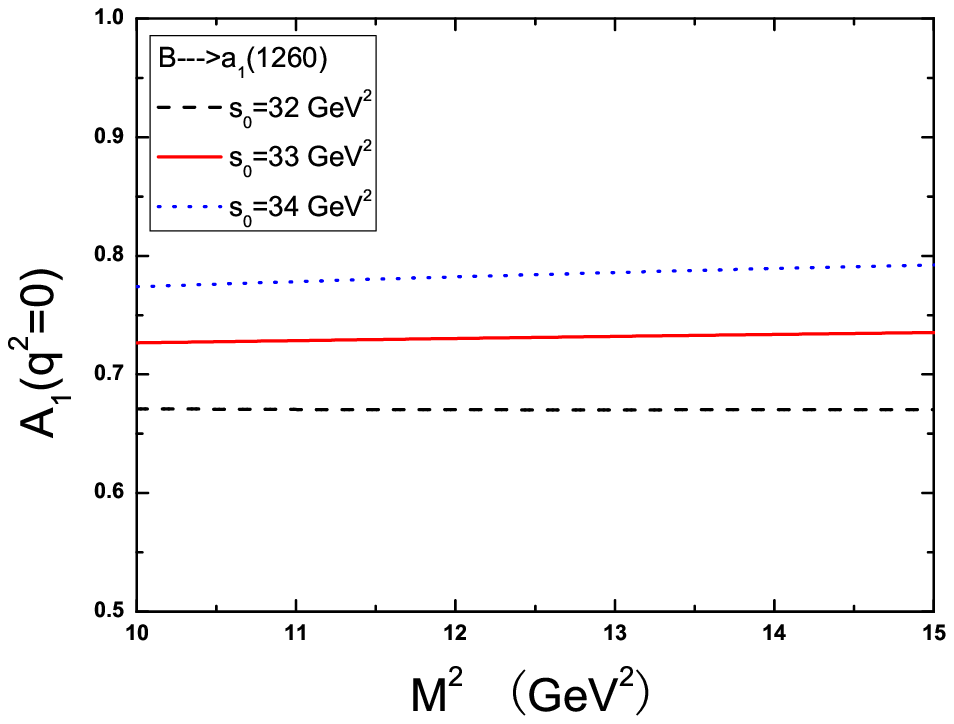}
\includegraphics[totalheight=5cm,width=6cm]{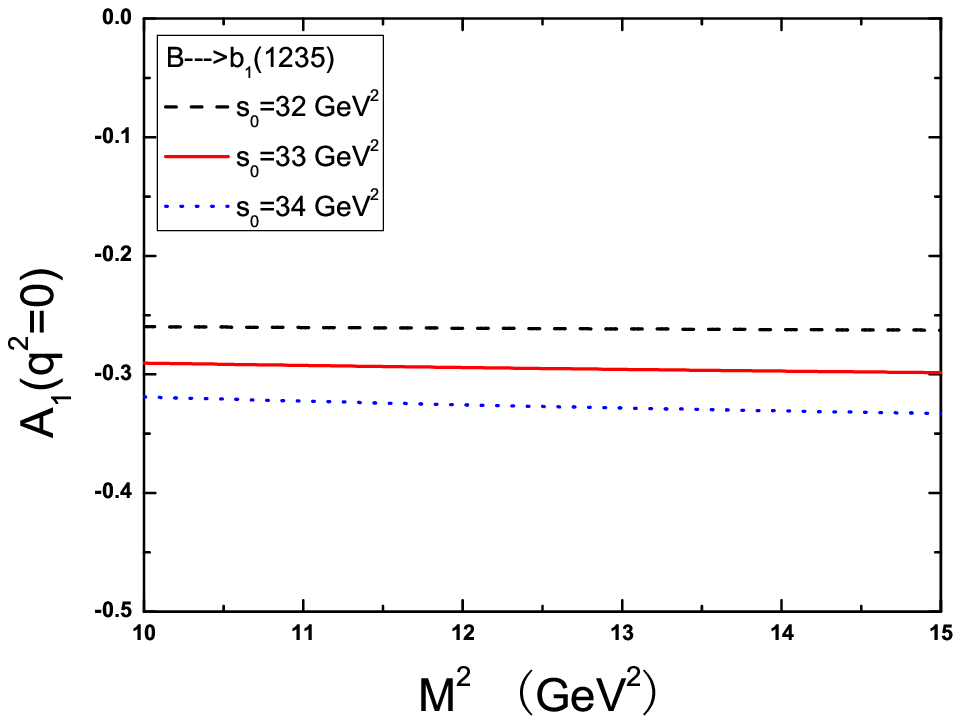}
\caption{The form-factor  $A_1(0)$ with variation of the  Borel
parameter $M^2$  at the energy scale $\mu=1.0 ~\mbox{GeV}$. The
threshold parameter $s_0=32,33,34 ~\mbox GeV^2$.} \label{figBA}
\end{figure}

The values of the form-factors $B\to a_1(1260)\, ,\, b_1(1235)$ at
zero momentum transfer are rather stable with variations of the
Borel parameter $M^2$. In Fig.\ref{figBA}, we present  numerical
results for the $A_1(0)$ with the central values of the input
parameters as an example.

The LCDAs of the  axial-vector mesons $^3P_1$ and $^1P_1$ have been
evaluated using the QCD sum rules \cite{Yang:2005gk,Yang:2007zt}.
Owing to the G-parity, the chiral-even two-particle  LCDAs of the
$^3P_1$ ($^1P_1$) mesons are symmetric (antisymmetric) under the
exchange of the quark and antiquark momentum fractions in the flavor
$SU(3)$ symmetry limit. For the chiral-odd LCDAs, the situation  is
versus. We show the numerical values of the LCDAs
$\Phi_{\perp}(u,\mu)$ of the axial-vector mesons $a_1(1260)$ and
$b_1(1235)$  at the energy scale $\mu=1.0~\mbox{GeV}$ explicitly in
Fig.\ref{figDA}. The integral interval
 in the sum rules is about $0.7 \sim 1$, and the decay
constants of the $a_1(1260)$ and $b_1(1235)$ mesons have the same
sign, therefore the form-factors $A_1, A_2, A_0, A$ for the  $B \to
a_1(1260)$, $b_1(1235)$ transitions have opposite sign, see Table
\ref{tab1}.  The uncertainties of the LCDAs $\Phi_\perp(u)$ and
constituent quark mass $m_b$ both result in errors for the
form-factors, which are shown as the first and second errors
respectively in Tab.\ref{tab1}.

\begin{table}
\begin{center}
\caption{The $B\to a_1(1260)\, , \, b_1(1235)$ form-factors  at zero
 momentum transfer, where
  the first  and second errors originate from the uncertainties of the LCDA $\Phi_\perp(u)$
 and the constituent quark mass $m_b$, respectively. In calculations, we have taken the values $M^2=12~\mbox{GeV}^2$ and
 $s_0=33~\mbox{GeV}^2$.}\label{tab1}
\begin{tabular}{|  c  |  c  |  c   |  c  |  c  |}\hline
 $B\to A$        & $A_1(0)$                    & $A_2(0)$                    & $A_0(0)$                     & $A(0)$ \\ \hline
 $B\to a_1(1260)$& $0.73{\pm 0.24}{\pm 0.11}$  & $0.41{\pm 0.14}{\pm 0.07}$  & $0.11{\pm 0.11}{\pm 0.01}$   & $0.41{\pm 0.14}{\pm 0.07}$  \\ \hline
 $B\to b_1(1235)$& $-0.29{\pm 0.09}{\pm 0.06}$ & $-0.17{\pm 0.06}{\pm 0.04}$ & $-0.05{\pm 0.05}{ \pm 0.05}$ & $-0.17{\pm 0.06}{\pm 0.04}$  \\ \hline
\end{tabular}
\end{center}
\end{table}

\begin{table}
\begin{center}
\caption{The $B\to a_1(1260)$ form-factors $A_1(0),A_2(0),A_0(0)$
and $A(0)$ from different theoretical approaches.}\label{tab2}
\begin{tabular}{|  c  |  c  |  c   |  c  |  c  |  c  |  c  |  c  |  c  |}\hline
            &CQM \cite{Deandrea:1998ww} &CLF \cite{Cheng:2003sm} &ISGW2 \cite{Scora:1995ty}  &QCDSR \cite{Aliev:1999mx} &LCSR \cite{Wang:2008bw}  &LCSR \cite{Yang:2008xw}  &pQCD \cite{Li:2009tx} &This work   \\ \hline
 $A_1(0)$   &2.10                       &0.59                    &0.87                       &0.68                      & 0.67                    & 0.60                    & 0.43                 & 0.73 \\ \hline
 $A_2(0)$   &0.21                       &0.11                    &-0.03                      &0.33                      & 0.31                    & 0.26                    & 0.13                 & 0.41 \\ \hline
 $A_0(0)$   &1.20                       &0.13                    &1.01                       &0.23                      & 0.29                    & 0.30                    & 0.34                 & 0.11 \\ \hline
 $A(0)$     &0.06                       &0.16                    &0.13                       &0.42                      & 0.41                    & 0.30                    & 0.26                 & 0.41 \\ \hline

\end{tabular}
\end{center}
\end{table}

We present the central values of the $B \to a_1(1260)$ form-factors
$A_1(0)$, $A_2(0)$, $A_0(0)$, $A(0)$ in Table {\ref{tab2}} compared
with the predictions from the CQM model \cite{Deandrea:1998ww}, CLF
approach \cite{Cheng:2003sm}, ISGW2 model \cite{Scora:1995ty}, QCDSR
\cite{Aliev:1999mx}, LCSR \cite{Wang:2008bw,Yang:2008xw}, and pQCD
\cite{Li:2009tx}. From the table, we can see that   the present
predictions are consistent with the ones from QCDSR
\cite{Aliev:1999mx} and LCSR \cite{Wang:2008bw,Yang:2008xw} except
for the $A_0(0)$, and differ from the values from other theoretical
approaches remarkably. It has been point out by K.C.Yang in
Ref.\cite{Yang:2008xw} that the higher twist effects might be
negligible, while we exclude  all contributions from the higher
twist LCDAs by using the chiral current in the correlation function.
In addition, the form-factors $A_1$, $A_2$, $A_0$, $A$
 are not independent, they are related with  the formulae like (\ref{eq:A-+}) and (\ref{eq:A+}).

\begin{figure}
\centering
\includegraphics[totalheight=5cm,width=6cm]{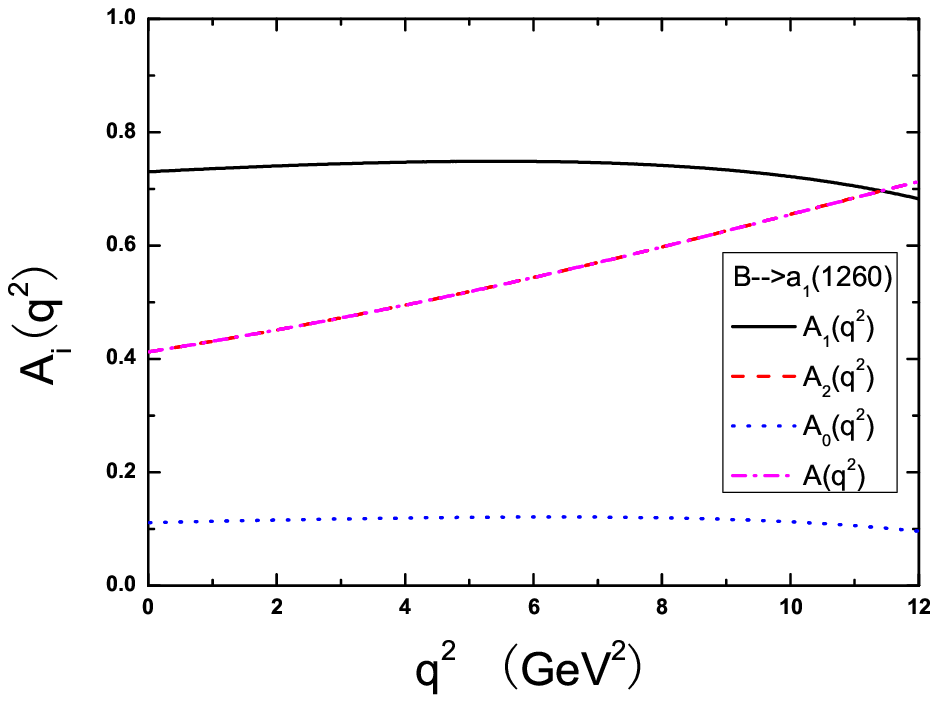}
\includegraphics[totalheight=5cm,width=6cm]{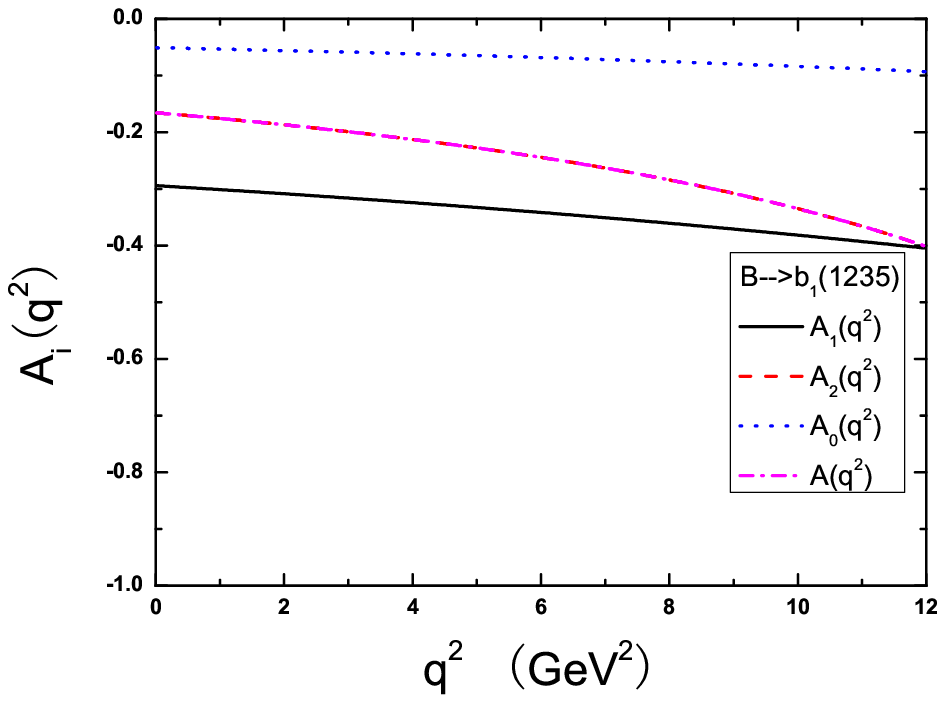}
\caption{The $B \to a_1(1260)$, $ b_1(1235)$ form-factors
$A_1(q^2)$, $A_2(q^2)$ and $A_0(q^2)$ with the  momentum transfer
$q^2$, where we have taken the values  $M^2=12 ~\mbox{GeV}^2$,
$s_0=33 ~\mbox{GeV}^2$ and $A_2=A$.}\label{fq}
\end{figure}

In Fig.\ref{fq}, we plot the $q^2$ dependence of the form-factors
$A_1(q^2)$, $A_2(q^2)$, $A_0(q^2)$, $A(q^2)$ for the transitions
$B\to a_1(1260)$,  $ b_1(1235)$ in the region $0\leq q^2<12
~\mbox{GeV}^2$, which is similar to the accessible  region $0\leq
q^2<10 ~\mbox{GeV}^2$ in the QCD sum rules \cite{Aliev:1999mx},
beyond that values the nonperturbative contributions  become large
and the operator product  expansion breaks down. The pole models are
merely suitable for describing those form-factors with momentum
transfers $q^2$ near the squared pole masses $m_{pole}^2$.  In the
present $B\to A$ case, the $m_{pole}^2$ are far away from their
kinematical regions, we do not extrapolate the  form-factors from
small $q^2$ to large ones with the pole models.

Now, we study the differential decay widths of the $B\to A$
semi-leptonic decays, which can be written as
\cite{Aliev:1999mx,Li:2009tx}
\begin{eqnarray}
&& \frac{d\Gamma_L(\bar B\to Al\bar\nu_l)}{dq^2} \label{eq:widthl}\\
&=&(\frac{q^2-m_l^2}{q^2})^2\frac{ {\sqrt{\lambda(m_{B}^2,m_A^2,q^2)}}
  G_F^2 V_{ub}^2} {384m_{B}^3\pi^3}
 \times \frac{1}{q^2} \left\{ 3 m_l^2 \lambda(m_{B}^2,m_A^2,q^2) V_0^2(q^2)+\right.\nonumber\\
 &&\;\;\times  \left.(m_l^2+2q^2) \left|\frac{1}{2m_A}  \left[
 (m_{B}^2-m_A^2-q^2)(m_{B}-m_A)V_1(q^2)-\frac{\lambda(m_{B}^2,m_A^2,q^2)}{m_{B}-m_A}V_2(q^2)\right]\right|^2
 \right\}, \nonumber\\
  &&\frac{d\Gamma_\pm(\bar B\to Al\bar\nu_l)}{dq^2}  \label{eq:widtht}\\
 &=&(\frac{q^2-m_l^2}{q^2})^2\frac{
 {\sqrt{\lambda(m_{B}^2,m_A^2,q^2)}} G_F^2V_{ub}^2}{384m_{B}^3\pi^3}
 \times   \nonumber\\
 &&\;\;\times \left\{ (m_l^2+2q^2) \lambda(m_{B}^2,m_A^2,q^2)\left|\frac{A(q^2)}{m_{B}-m_A}\mp
 \frac{(m_{B}-m_A)V_1(q^2)}{\sqrt{\lambda(m_{B}^2,m_A^2,q^2)}}\right|^2
 \right\},\nonumber
\end{eqnarray}
where $\lambda(m_B^2,m_A^2,q^2)=(m_B^2+m_A^2-q^2)^2-4m_B^2m_A^2$,
and $L,+,-$ denote the helicities of the  axial-vector mesons.

\begin{figure}
\centering
\includegraphics[totalheight=5cm,width=6cm]{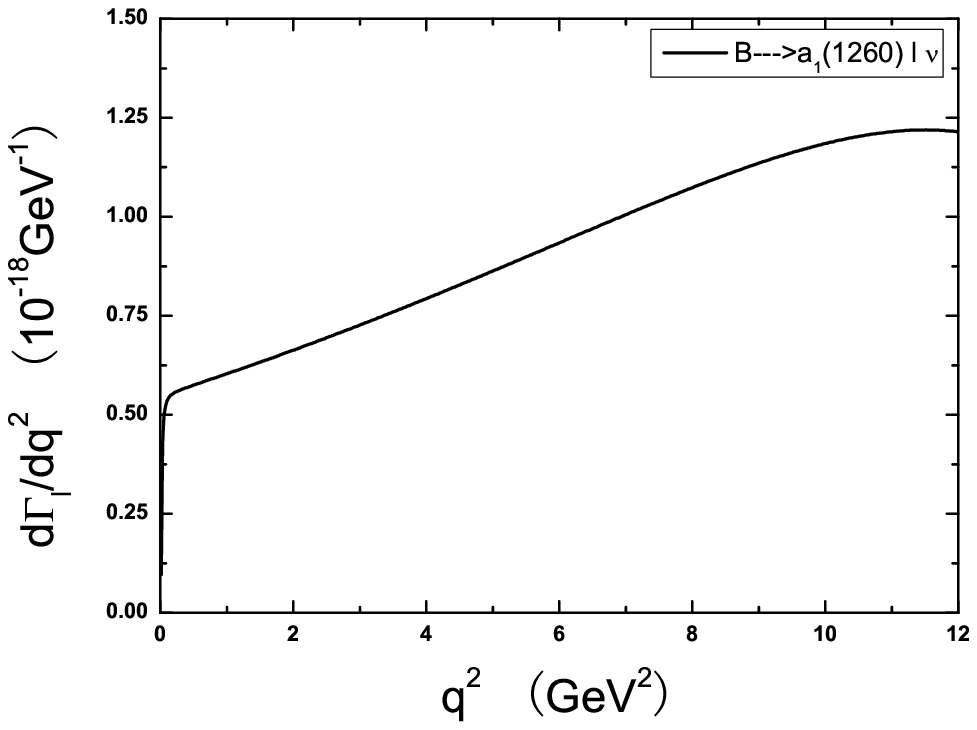}
\includegraphics[totalheight=5cm,width=6cm]{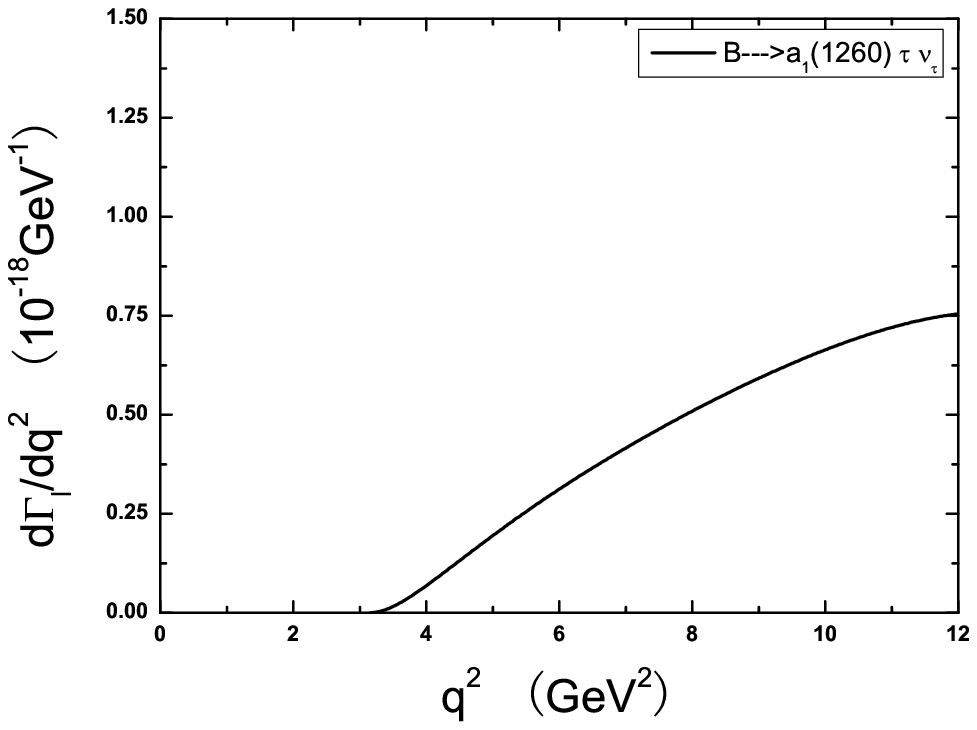}
\includegraphics[totalheight=5cm,width=6cm]{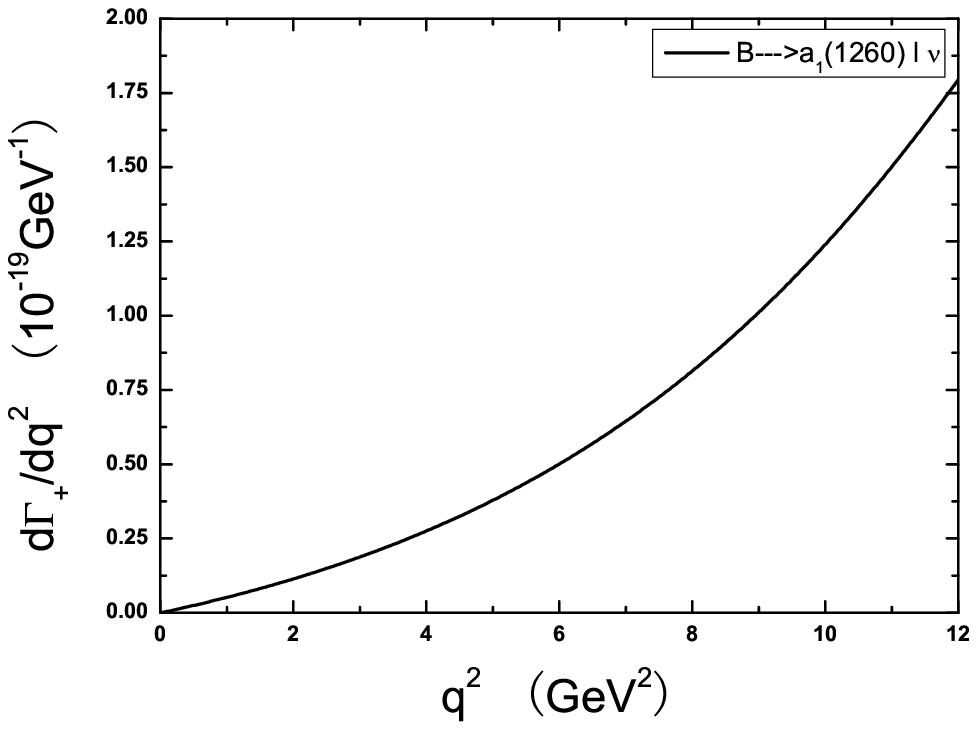}
\includegraphics[totalheight=5cm,width=6cm]{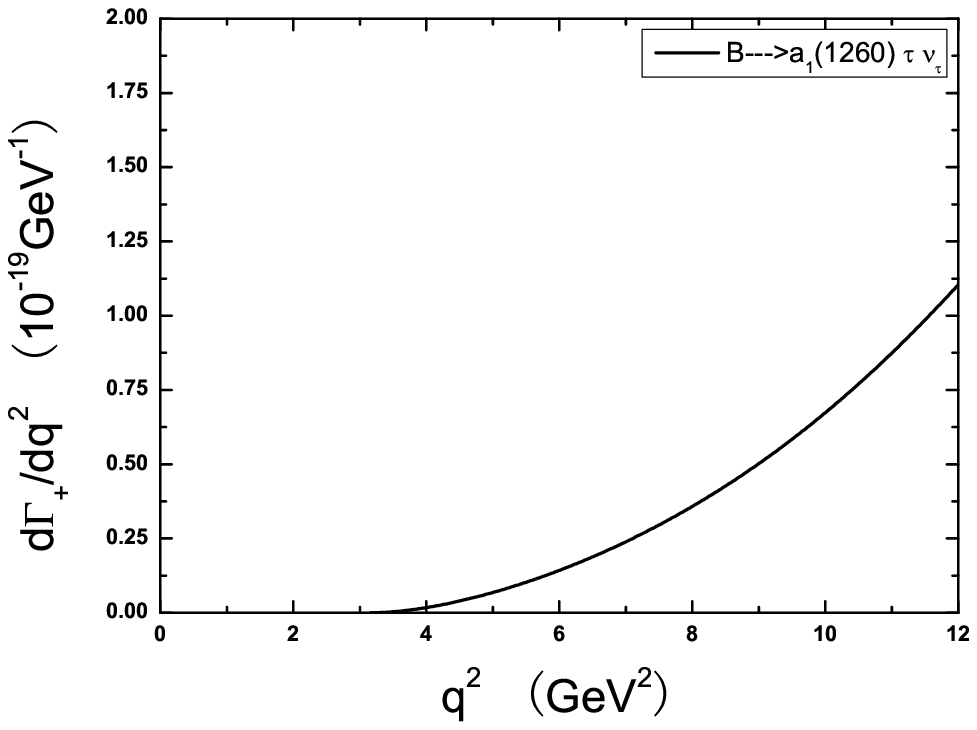}
\includegraphics[totalheight=5cm,width=6cm]{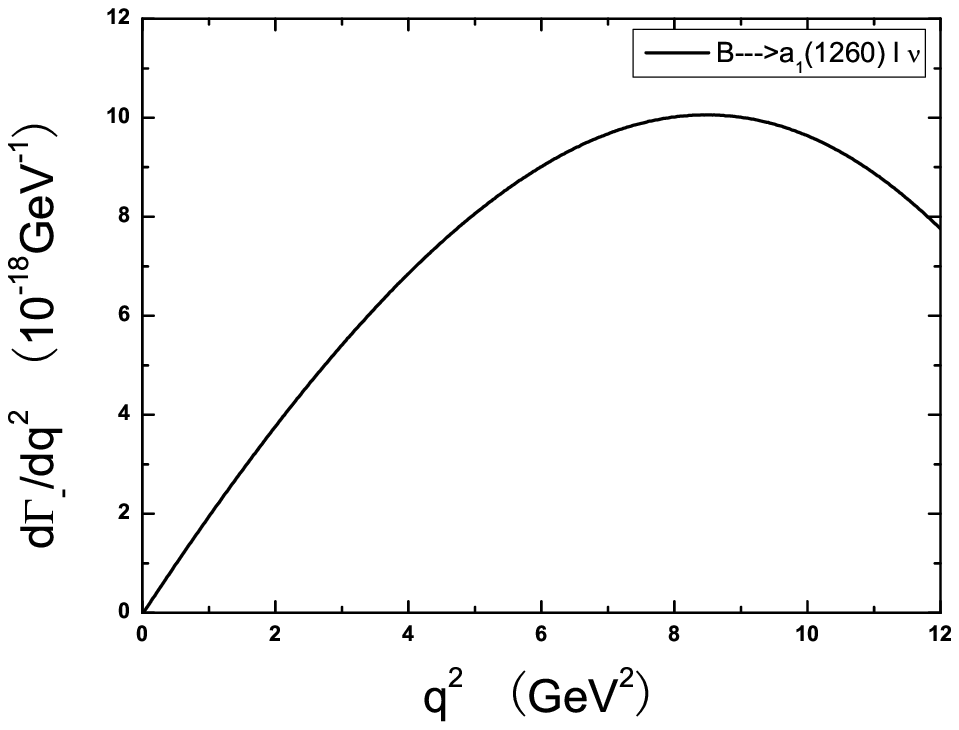}
\includegraphics[totalheight=5cm,width=6cm]{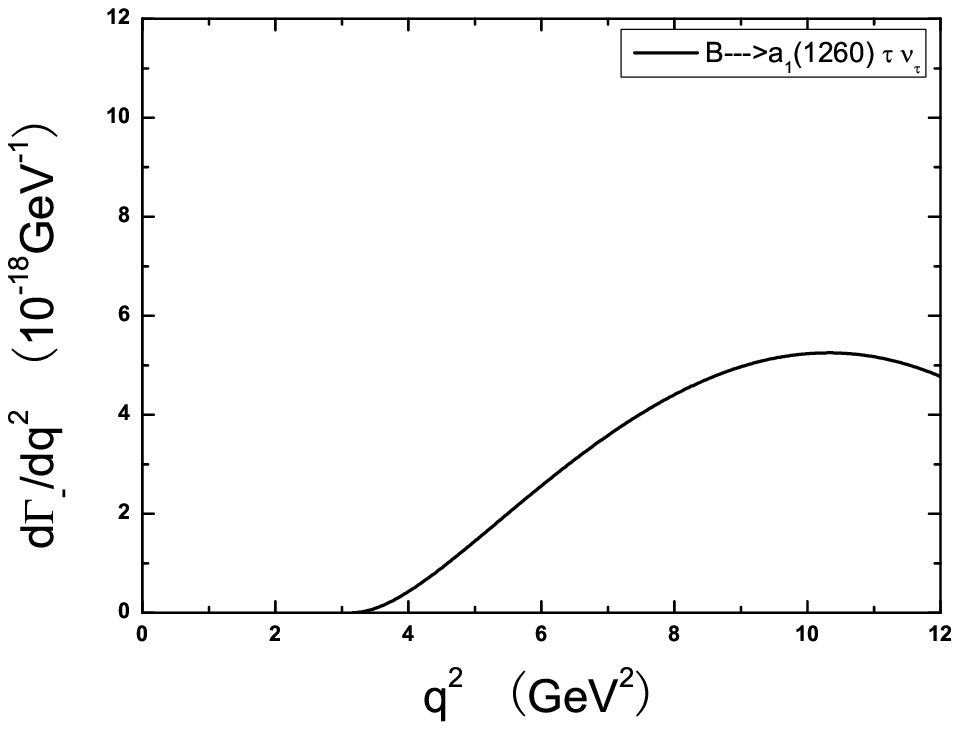}
\caption{Differential decay widths of the $B\to
a_1(1260)l\bar{\nu}_l$  as functions of $q^2$. Here $l=e,\mu$ in the
left diagram.}\label{widthBAa1}
\end{figure}

\begin{figure}
\centering
\includegraphics[totalheight=5cm,width=6cm]{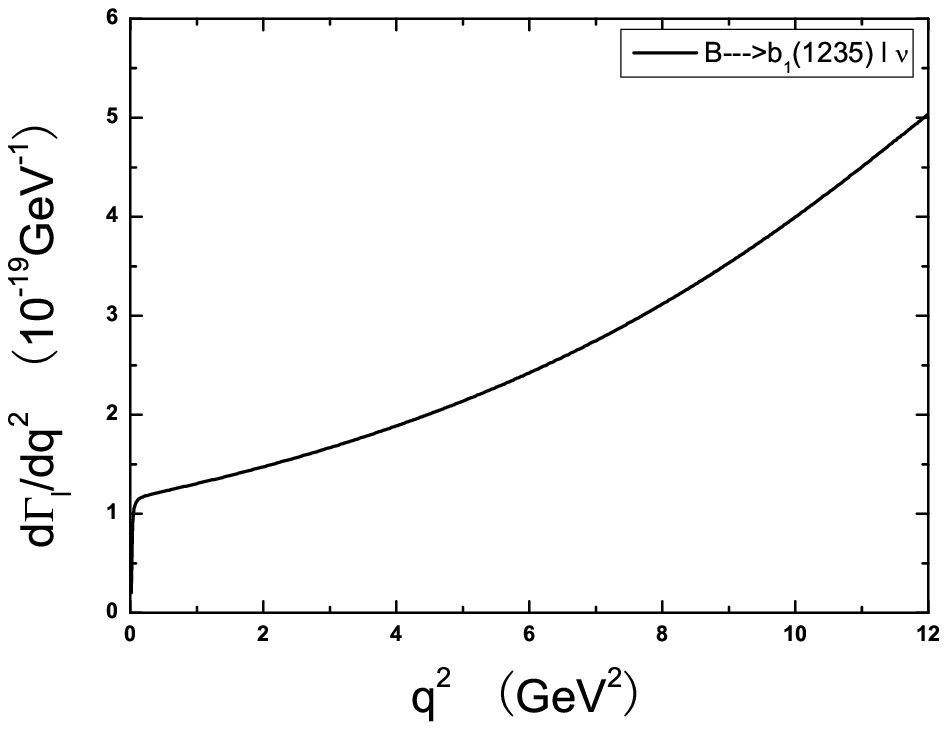}
\includegraphics[totalheight=5cm,width=6cm]{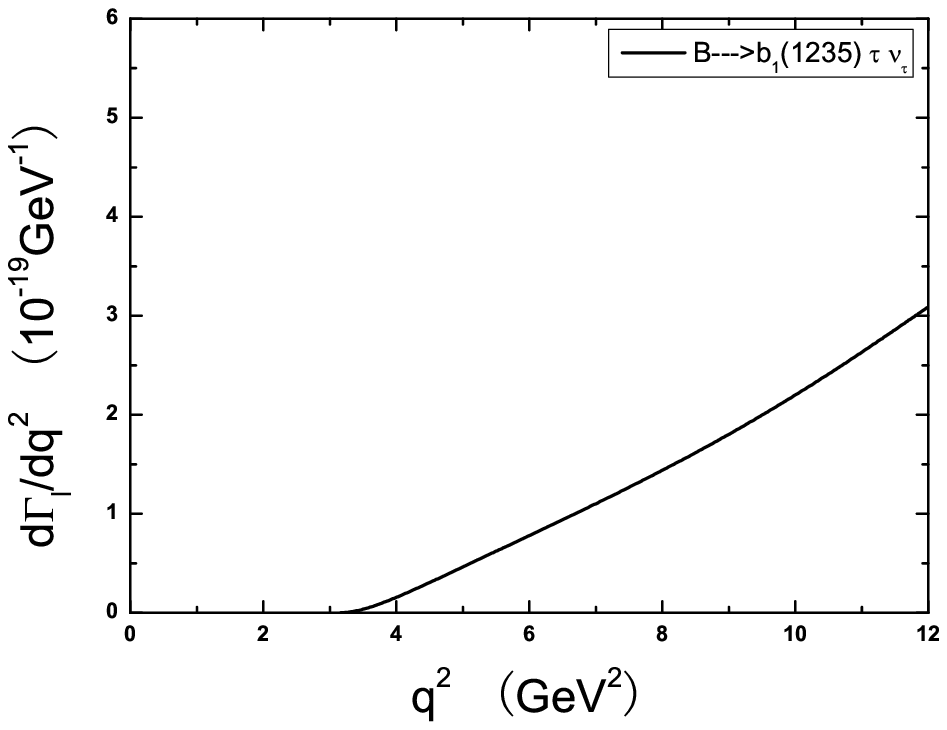}
\includegraphics[totalheight=5cm,width=6cm]{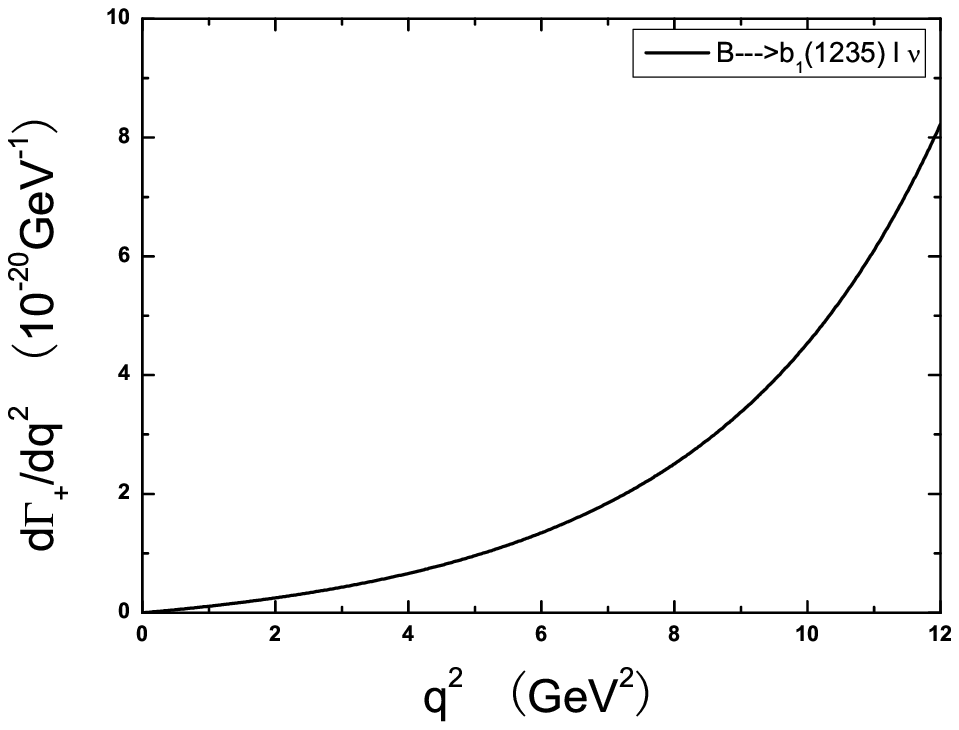}
\includegraphics[totalheight=5cm,width=6cm]{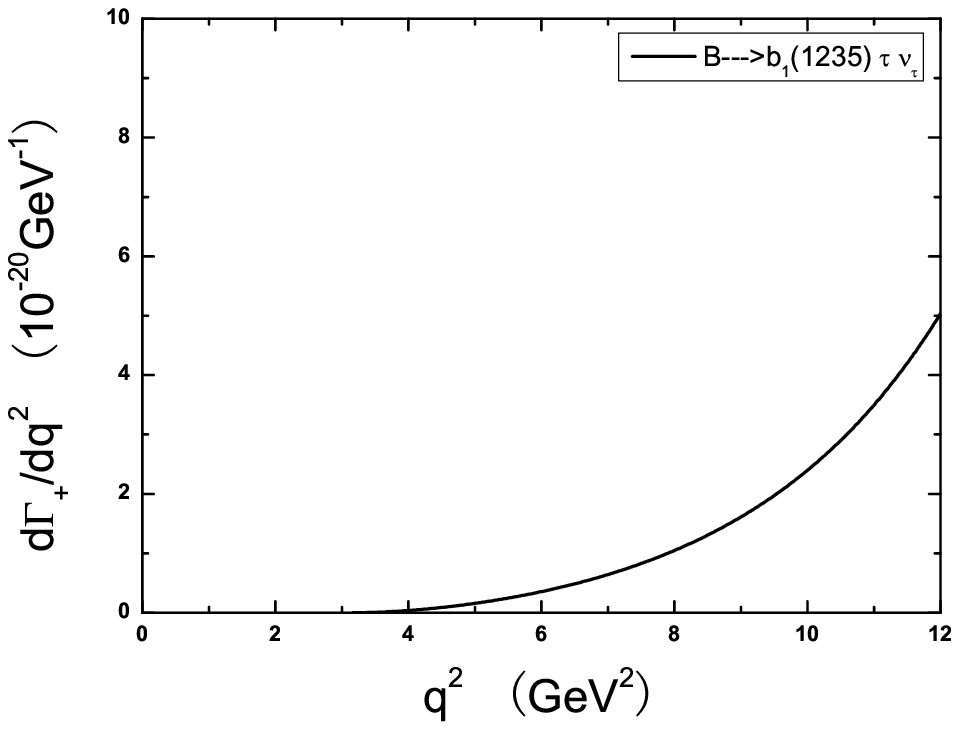}
\includegraphics[totalheight=5cm,width=6cm]{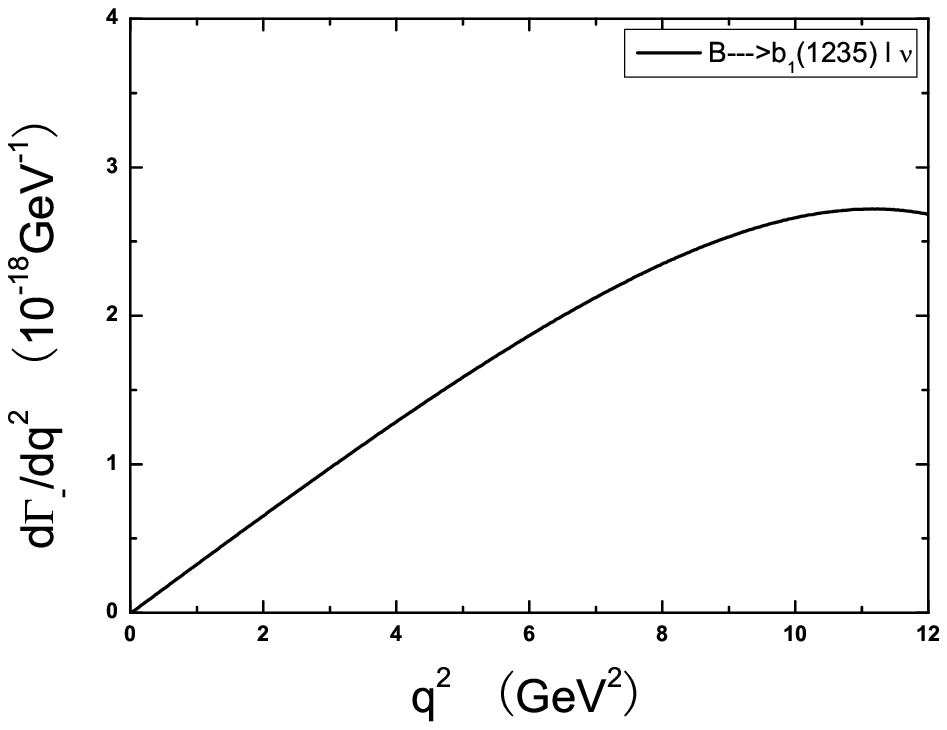}
\includegraphics[totalheight=5cm,width=6cm]{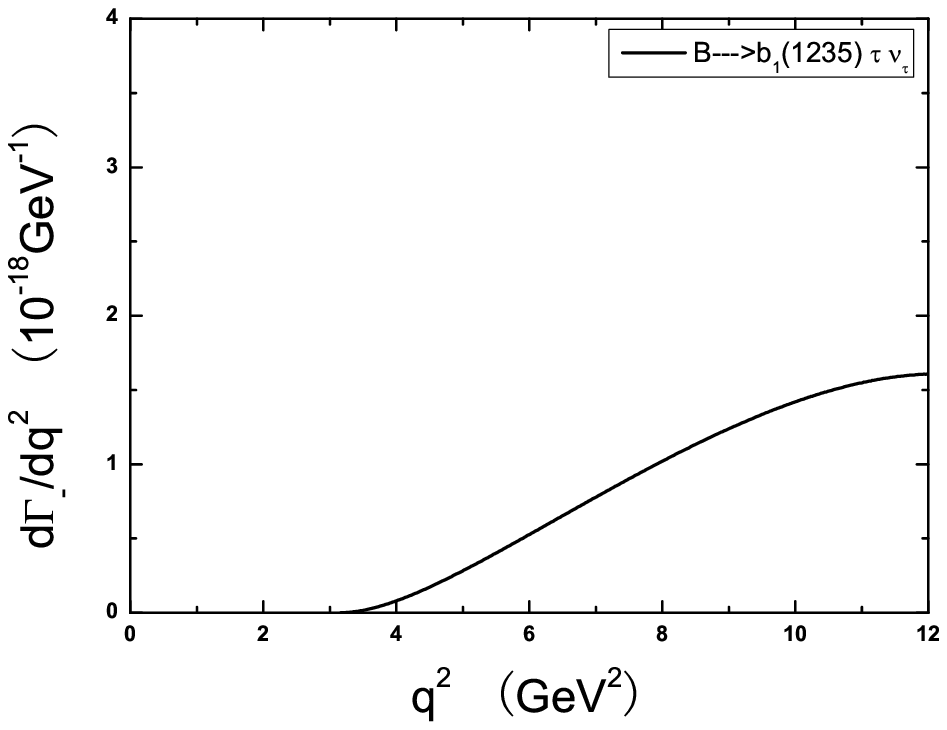}
\caption{Differential decay widths of the  $B\to
b_1(1235)l\bar{\nu}_l$  as functions of $q^2$. Here $l=e,\mu$ in the
left diagram.}\label{widthBAb1}
\end{figure}

 We plot  the differential decays widths  of  the   $B\to
a_1(1260)l \bar{\nu}_l$, $ b_1(1235) l \bar{\nu}_l$ in the effective
regions $m_l^2 \leq q^2\leq (m_B-m_A)^2$ in
Figs.\ref{widthBAa1}-\ref{widthBAb1},  where we  take
$m_e=m_{\mu}=0$. We can integrate the differential decay widths over
the variable $q^2$, and obtain the decay widths, which satisfy the
relation $\Gamma_- > \Gamma_L \gg \Gamma_+$, and are consistent with
the results of Ref.\cite{Yang:2008xw}.

\section{Summary and discussion}

In this article, we calculate the $B\to a_1(1260)$, $ b_1(1235) $
form-factors in the accessible  region  $0\leq q^2 < 12
~\mbox{GeV}^2$ with the light-cone QCD sum rules  at the leading
order approximation, then study the differential decay widths and
decay widths of the  semi-leptonic decays $B\to a_1(1260) l
\bar{\nu}_l$, $b_1(1235) l \bar{\nu}_l$. \\

(1) In this paper, we choose the chiral current to interpolate the
$B$-meson, and observe that only the leading-twist LCDAs of the
axial-vector mesons contribute to the form-factors after taking
account of the transversely polarization of the axial-vector mesons.
We avoid contributions from the twist-3 LCDAs which have the most
uncertainty in the form-factors by using the chiral current. The
uncertainties originate from the LCDAs are reduced remarkably.\\

(2) Owing to the G-parity of the axial-vector mesons $^3P_1$ and
$^1P_1$, the form-factors of the $B \to a_1(1260)$, $ b_1(1235)$
transitions have opposite sign. There exist relations among the
$B\to A$ transition form-factors which are in accordance with the
prediction of the soft collinear effective theory
\cite{Bauer:2000yr}.\\

(3) The present predictions of the differential decay widths and
decay widths of the  semi-leptonic decays $B\to a_1(1260) l
\bar{\nu}_l$, $b_1(1235) l \bar{\nu}_l$ can be confronted with the
experimental data at the KEK-B and LHCb in the coming future.  If
the perturbative $\mathcal {O}(\alpha_s)$ corrections  are taken
into account, the predictions may be improved,   however, the
improvements are not expected to be large considering  the
corresponding calculations of the $B\to V$ form-factors.\\

{\bf Acknowledgments}: Authors would like to thank Dr.Fen Zuo for
helpful discussions. This  work is supported by National Natural
Science Foundation of China, Grant Numbers 10735080, 10805082,
10675098, 11075053,  and the Fundamental Research Funds for the
Central Universities.

\end{document}